\documentclass[a4paper,cite]{article}
\usepackage{latexsym}

\usepackage[latin1]{inputenc}

\usepackage{graphicx}
\usepackage{color,pst-plot}

\usepackage{amsmath}
\usepackage {amsfonts,amssymb,amstext}

\def\bel{\begin{equation}\label}

\newcommand{\be}{\begin{equation}}
\newcommand{\ee}{\end{equation}}

\newcommand{\beq}{\begin{equation}}
\newcommand{\eq}{\end{equation}}

\newcommand{\bea}{\begin{eqnarray}}
\newcommand{\eea}{\end{eqnarray}}

\newcommand{\noi}{\noindent}

\DeclareMathAlphabet{\eusm}{U}{}{}{}
\SetMathAlphabet\eusm{normal}{U}{eus}{m}{n}
\SetMathAlphabet\eusm{bold}{U}{eus}{b}{n}
\DeclareMathAlphabet{\mathpzc}{OT1}{pzc}{m}{it}

\input epsf

\begin{document}

\begin{titlepage}

\begin{flushright}
CPHT-RR032.0508\\
LPT-08-52\\
arXiv:0805.4346\\

\end{flushright}

\vspace*{0.2cm}
\begin{center}
{\Large {\bf On AdS/QCD correspondence and  \\
the partonic picture of deep inelastic scattering 
}}\\[2 cm]

{\bf B. Pire}~$^a$,  {\bf C. Roiesnel}~$^a$, {\bf  L. Szymanowski}~$^{ b}${\bf and S. Wallon}~$^{ c}$\\[1cm]

$^a$  {\it Centre  de Physique Th{\'e}orique, \'Ecole Polytechnique, CNRS,
   91128 Palaiseau, France}\\[0.5cm]
$^b$ {\it Soltan Institute for Nuclear Studies, Warsaw, Poland}\\[0.5cm]
$^c$ {\it LPT, Université d'Orsay, CNRS, 91404 Orsay, France}
\end{center}

\vspace*{3.0cm}

\vspace*{3.0cm}
\begin{abstract}
  We critically examine the question of scaling of the Deep Inelastic
  Scattering process in the medium Bjorken $x$ region on a scalar
  boson in the framework of the AdS/QCD correspondence.  To get the
  right polarization structure of the forward electroproduction
  amplitude, we show that one needs to add (at least) the scalar to
  scalar and scalar to vector hadronic amplitudes.  This
  illustrates 
  how the partonic picture may emerge from a simple scenario based on
  the AdS/QCD correspondence, provided one allows the conformal
  dimension of the hadronic field to equal $1$ and use the concept of
  "hadron - parton duality" .
  \end{abstract}

\end{titlepage}

\section{\normalsize Introduction}
The near conformal symmetry of QCD has driven a number of theoretical
attempts named as the AdS/QCD correspondence
to obtain some information about scattering amplitudes of processes at
high energies.  The seminal paper of Polchinski and Strassler
\cite{PS1} showed how one may hope to recover some scaling laws in the
realm of hard hadronic reactions. Exclusive amplitudes have also been
studied in a number of cases \cite{BdT, Rad} and a partonic
interpretation has been proposed by Brodsky and de Teramond \cite{BdT}
in the framework of the light-front dynamics.  Various aspects of deep
inelastic scattering have also been studied~\cite{Karch,BBB,others,CC}.

In this paper we critically scrutinize the simplest case, analysed
also in \cite{PS1}, where the partonic structure emerges from an OPE
analysis of a scattering amplitude, namely the imaginary part of the
forward amplitude for the process
$$\gamma^* H \to \gamma^* H$$ where $H$ is a  hadron, which is related to the
 total cross section for the deep inelastic electroproduction process
$$\gamma^* H \to X $$  by the usual optical theorem. 

To begin with, let us outline the well-known two facts that seem to us
essential to the understanding of the deeply inelastic scattering
(DIS) process on a hadron in terms of the underlying QCD process
where partons, namely quarks in the lowest order, respond to the
electromagnetic current.
\begin{itemize}
\item The amplitude scales as $Q^0$ (up to logarithmic modifications)
  at large $Q^2 $ and fixed Bjorken variable ($x_{Bj}= Q^2/s $),
  where $Q^2 = - q^2$ is the virtuality of the photon and $s$ the
  squared energy of the process, and this scaling behaviour is the
  signal that the electromagnetic current scatters on pointlike
  particles inside the hadron.
\item The leading amplitude corresponds to the case of a
  transversally polarized virtual photon, which is the signal that
  these pointlike constituents are fermions.
\end{itemize}
Let us stress that these features do not depend on the spin of the
hadronic target, so that we feel free to restrict our study to the
technically simplest case of a scalar boson (e.g.  a $f^{0}$ meson,
for definiteness).
 
Our aim is to explore whether we can get these two features from an
analysis of DIS in the framework of the AdS/QCD correspondence, in
its simplest version as defined in the paper of Polchinski and
Strassler~\cite{PS2}, {\em i.e.} the hard-wall model
 (our discussion actually does not depend on this choice: see Appendix for 
the case of soft-wall models).
The conclusions of Ref. \cite{PS2} show that there is no obvious
answer to this question, and one may even wonder whether the results
of this approach establish a bridge between the partonic and AdS/QCD
descriptions of DIS.
 
The plan of the paper is as follows. In section 2, we briefly remind
the reader of the basics of the AdS/QCD correspondence strategy to
analyse the amplitude of the process $ \gamma^* H \to X$ where $X$ is
a massive state that corresponds to the final state of the DIS
process. Taking $X$ as a scalar object, we recover the results of
Ref. \cite{PS2} and discuss their virtues and defects with respect to
a partonic interpretation in the kinematical domain known as the
Bjorken scaling region. Motivated by the defects discovered, we
calculate in section 3 the corresponding process where $X$ is now a
spin 1 object. In section 4, we argue that the physically sensible
amplitude is a weighted sum of the spin 0 and spin 1 amplitudes and
show that indeed we can recover the needed features of the DIS
amplitude, provided we allow the conformal dimension of the hadronic
field to equal $1$.

\section{\normalsize DIS on a scalar target with a scalar intermediate
  state}
Let us recall the results obtained in Ref.~\cite{PS2} for the
calculation of the amplitude of the process $\gamma^*(q) \; S \;\to \;
X,$ where $S$ and $X$ are scalar states.  The basic quantities are a
massless vector field $A^m(x,z)$ and massive complex scalar fields
$\Phi(x,z)$ which are treated as free modes propagating in a
five-dimensional Anti-de Sitter space (AdS$_5$) with curvature $R$. We
shall use the Poincar\'e coordinates with metric $g_{mn}$ defined by
\begin{align}
  ds^2 = \frac{R^2}{z^2}\left(\eta_{\mu\nu}dx^{\mu}dx^{\nu} + dz^2\right)\ \,,
\end{align}
where $\eta_{\mu\nu}=(-,+,+,+)$ is the Minkowski metric.

The massless vector field $A^m(x,z)$ is identified with the
electromagnetic field and obeys the Maxwell equations in AdS$_5$ with
the boundary conditions
\begin{equation} 
\label{boundaryA} \lim_{z \to 0} A_\mu(x,z) =
A_\mu(x)|_{\rm 4d}\ = n_\mu \, e^{i q \cdot x} \,,\qquad
\lim_{z \to 0} A_z(x,z) = 0\ \,,  
\end{equation}
where $n_{\mu}$ is an arbitrary polarization vector. With the
Lorentz-like gauge condition $\eta_{\mu\nu}\partial^\mu A^{\nu} + z \,
\partial_z \left(\frac{1}{z} A_z \right)=0 \,$,
the solution for a spacelike photon, $Q^2=q\cdot q > 0$, reads: 
\bea
\label{solAmueps+1}
A_{\mu}(x,z)=Q \, z \, K_1(Q \, z) \, e^{i q \cdot x} n_\mu\ , ~~~~~~
A_z(x,z)=i(q \cdot n)\,z \, K_0(Q \, z) \, e^{i q \cdot x}  \,.
\eea

The electromagnetic current is written as
\begin{equation}
  J_m(x,z) = i\sum_{X}\phi_0\partial_m\phi^{\star}_X 
  - \phi^{\star}_X\partial_m\phi_0 \;,
\end{equation}
where the scalar initial field $\phi_0(x,z)$ and final field
$\phi_X(x,z)$ are normalizable classical solutions of the Laplacian in
Anti-de Sitter space, which is a representation of the quadratic
Casimir operator of the isometry group $SO(4,2)$ of AdS$_5$. These
solutions belong to the same irreducible representation of $SO(4,2)$
labelled by the conformal dimension $\Delta_0$\footnote{Indeed 4d
  gauge invariance implies that $\Delta_X=\Delta_0$.},
\begin{gather}
  \phi_X(x,z) = c_Xe^{ip_X\cdot x}z^2J_{\Delta_0-2}(s^{1/2}z)\,,\qquad
    \Delta_0(\Delta_0-4) = m_5^2R^2\,,\quad p_X^2=-s<0 \ \,,\\
    \lim_{z\rightarrow 0}\phi_X(x,z) \propto z^{\Delta_0}e^{ip_X\cdot x}\ \,.
\end{gather}
The constant $c_X$ is fixed by imposing a Dirichlet boundary condition
at $z_{\infty}=1/\Lambda$ and by normalizing to unity the charge density
$J^0(x,z)$ of the field $\phi_X$ over each time slice,
\begin{gather}
  \label{Chard}
  i\int_0^{z_{\infty}} dz\,\frac{R^3}{z^3}
    (\phi_X\partial_0\phi^{\star}_X - \phi^{\star}_X\partial_0\phi_X) 
    = 2\,E\quad\Longrightarrow\quad
  c_X = \frac{\sqrt{2}\Lambda}{\sqrt{R^3}}J^{-1}_{\Delta_0-1}
  \left(\frac{\sqrt{s}}{\Lambda}\right) 
  \underset{\Lambda\rightarrow 0}{\sim}
  \frac{\Lambda^{1/2}s^{1/4}}{\sqrt{R^3}} \\
  \sqrt{s} = m_n = \zeta_{\Delta_0-2,n}\Lambda\,,\qquad 
  J_{\Delta_0-2}(\zeta_{\Delta_0-2,n}) = 0\ \,.
\end{gather}
The normalization integral is defined for $\Delta_0> 1$.

The covariant interaction with the electromagnetic field $A_m(x,z)$ reads, to
first order in the electromagnetic coupling $e$,
\begin{equation}
  S_{int} = e\int d^4x dz \sqrt{-g}~ g^{mn}A_m(x,z)J_n(x,z)\ \,.
\end{equation}
In the low energy limit for the initial scalar field, 
\begin{align}
  \label{lowE}
  \phi_0(x,z) \underset{s_0\rightarrow 0}{\sim} 
  c_0z^{\Delta_0}e^{ip_0\cdot x}\,,\qquad
  c_0\propto \frac{\Lambda^{\Delta_0-1}}{\sqrt{R^3}}\ \,,
\end{align}
the $\gamma ^{*} S \to X$ amplitude is proportional to
\begin{align}
  M^0_{\mu} = Q^{-\Delta_0}A(x) \left(p_{0\mu}+\frac{1}{2x}q_{\mu}\right)
  \,,\quad  A(x) = 2^{\Delta_0}\Gamma(\Delta_0)
  x^{\frac{\Delta_0}{2}+1}(1-x)^{\frac{\Delta_0}{2}-1}
  \,,\quad x = -\frac{Q^2}{2p_0\cdot q}\ \,.
\end{align}
Plugging in the normalization constants one gets for the
electromagnetic hadronic tensor,
\begin{gather}
  W^0_{\mu\nu} = \frac{1}{2}(c_0c_XR^3)^2
  \sum_{n}\delta(s-m_n^2) Q^{-2\Delta_0}
  \left(p_{0\mu}+\frac{1}{2x}q_{\mu}\right)
  \left(p_{0\nu}+\frac{1}{2x}q_{\nu}\right)A^2(x) \; .
\end{gather}
Hence, performing the sum over $n$, with a density of states
\begin{align}
\label{Dhard}
\sum_n\delta(s-m_n^2) \approx \int dn \delta(n^2\pi^2\Lambda^2-s) = 
(2\pi\sqrt{s}\Lambda)^{-1}\,,
\end{align}
the usual structure functions
$F_1(x,Q^2)$ and $F_2(x,Q^2)$ can be identified as
\begin{align}
\label{PSF1F2}
 F_1(x,Q^2) = 0 \,,\qquad
  F_2(x,Q^2) = \frac{c}{4\pi}\left(\frac{\Lambda}{Q}\right)^{2\Delta_0-2}
  \frac{1}{2x}A^2(x)\; .
\end{align}
where $c$ is some dimensionless constant. One could wonder whether the
$Q^2$ behavior depends upon the specific mechanism of conformal
symmetry breaking of the hard-wall model. In the Apppendix we show
that this $Q^2$-dependence is quite general since one obtains the same
result in the soft-wall models where one introduces background dilaton
fields \cite{Karch}.
 
\vskip 0.2cm
\noindent
We can now address the two questions emphasized in Section 1. 
\begin{itemize}
\item Firstly, the $Q^2$ behaviour is controlled by the dimension
  $\Delta_{0}$ attached to the incoming field. 
  If one identifies this dimension to the conformal dimension of a
  hadron seen as a bound state of elementary quarks and gluons (namely
  $\Delta_{0} = 2$ for a meson)
  as proposed in the analysis of exclusive reactions~\cite{PS1,
    BdT}, we completely miss the scaling behaviour of DIS structure
  functions which is central to the parton picture~\cite{Feynman} and
  to the recognition of QCD as the theory of strong interactions able
  to understand the experimental data.  In order to make contact with
  reality, one is thus lead to consider the case where
    $\Delta_{0} = 1$, which may be interpreted as the recognition
  that the incoming hadron fluctuates to an elementary field before
  scattering with the virtual photon. Most interestingly, it turns out
  that this value of $\Delta_{0}=1$ coincides with the unitarity bound
  on the dimension of scalar operator in four dimensions
 \cite{KW}. We thus propose to take seriously 
 $\Delta_{0} = 1$ as the starting point of a partonic interpretation of the AdS
 calculation of DIS structure functions.

\item The fact that $F_{1}$ vanishes shows that the tensorial
  structure of the result is completely at odds with a sensible
  partonic interpretation.
 
 \end{itemize}
 We thus conclude that although a scaling behaviour may be marginally
 recognized in the results of the AdS/QCD calculation the tensorial
 structure of the amplitude prevents us from being able to make
 contact with the partonic interpretation.  This is by no ways a
 surprise, since the spin structure is indeed very poor when one
 considers a scalar to scalar amplitude and indeed, the restriction to
 a scalar meson in the final state is completely unjustified since the
 kinematics considered emphasize large (4-dimensional) invariant
 masses $M_{X}^2 = s$.
 
 We thus advocate that it is very important to consider the case of
 higher spin final states and, as a concrete example, calculate the corresponding observables
 for a spin 1 final state.
 
 \section{\normalsize DIS on a scalar target with a vector
   intermediate state}
 The missing ingredient for a spin 1 final state is the solution
 $V_m(x,z)$ of the free massive Maxwell equations in AdS$_5$,
\begin{gather}
  G_{mn} = \partial_mV_n - \partial_nV_m\ \,, \\
  \left(D_mG\right)^{mn}\equiv 
  \frac{1}{\sqrt{-g}}\partial_m\left(\sqrt{-g}G^{mn}\right) = m_V^2V^n
  \,,\qquad\forall n\ \,.
\end{gather}
We look for a solution in the radial gauge such that
\begin{align}
  V_{\mu}(x,z) = C e_{\mu}e^{ik\cdot x}V(z)\,,\qquad V_z = 0\,,\quad 
  e\cdot k=0\,,\quad e^2=1\ \,,
\end{align}
and which belongs to an irreducible representation of $SO(4,2)$
labelled by the conformal dimension $\Delta_V$ and characterized by
the asymptotic behavior
\begin{align}
  \lim_{z\rightarrow 0} V_{\mu}(x,z) \propto z^{\Delta_V}e^{ik\cdot x}\ \,.
\end{align}
The solution with the right boundary conditions reads
\begin{align}
  \label{radial}
  V_{\mu}(x,z) = C_{\Delta_V}e_{\mu}e^{ik\cdot x}zJ_{\Delta_V-1}(\sqrt{s}z)
  \,,\qquad\Delta_V(\Delta_V-2) = (m_VR)^2\,,\quad k^2=-s<0\ \,.
\end{align}
The normalization of charged vector fields can be defined as for
charged scalar fields by imposing a Dirichlet boundary condition at
$z_{\infty}=1/\Lambda$ and by normalizing to unity the charge density
$J^0(x,z)$ of the field $V_{\mu}$ over each time slice,
\begin{gather}
  i\int_0^{z_{\infty}} dz\,\frac{R}{z}
    (V_{\nu}G^{\star\,0\nu} - V_{\nu}^{\star}G^{0\nu}) 
    = 2\,E\quad\Longrightarrow\quad
  C_{\Delta_V} = \frac{\sqrt{2}\Lambda}{\sqrt{R}}J^{-1}_{\Delta_V}
  \left(\frac{\sqrt{s}}{\Lambda}\right) 
  \underset{\Lambda\rightarrow 0}{\sim}
  \frac{\Lambda^{1/2}s^{1/4}}{\sqrt{R}} \\
  \sqrt{s} = m_{V,n} = \zeta_{\Delta_V-1,n}\Lambda\,,\qquad 
  J_{\Delta_V-1}(\zeta_{\Delta_V-1,n}) = 0\ \,.
\end{gather}
The normalization integral is defined for $\Delta_V>0$.
We note that the $s$ dependence of the normalization constant and of
the sum over the intermediate states are the same for a massive scalar
field and for a massive vector field (in this so-called ``hard-wall''
model).

The interaction between the electromagnetic field $A_m$, a scalar
field $S$ and a vector field $V_m$ can be described to first order in
the electromagnetic coupling $e$ by an effective gauge-invariant
lagrangian with non-minimal coupling of the following form
\begin{align}
  \mathcal{L}_{V\gamma S} = e\frac{g_{V\gamma S}}{m_V}
  \left(\partial^iV^j\partial_iA_j - \partial^iV^j\partial_jA_i\right)S\ \,.
\end{align}
The covariant interaction in the Anti-de Sitter space AdS$_5$ reads
\begin{align}
  S_{int} = ie\frac{g_{V\gamma S}}{m_V}\int d^4x dz\sqrt{-g}\,\Phi_0\,
  \left(g^{mp}g^{nq}\partial_pV^{\star}_qF_{mn}\right)
\end{align}
where $F_{mn}$ is the electromagnetic tensor.
In the low energy limit \eqref{lowE} for the initial scalar field, and
with the massive vector field in the radial gauge \eqref{radial}, we
get (with implicit summation over mute indices understood with the
four-dimensional Minkowski metric)
\begin{align*}
  S_{int} &= ie\frac{g_{V\gamma S}}{m_V}c_0\int d^4x dz\frac{R}{z} 
  z^{\Delta_0} 
  \left(\partial^{\mu}V^{\nu\star}F_{\mu\nu}+\partial_zV^{\mu\star}F_{z\mu}\right)
  \\ &= ie\frac{g_{V\gamma S}}{m_V}c_0c_VR\,\delta^{(4)}(k-p_0-q)
  \int dz\,z^{\Delta_0-1} \\
  &\quad\times
  \biggl(Q\left((k\cdot q)(n\cdot e)-(k\cdot n)(q\cdot e)\right)
    z^2J_{\Delta_V-1}(\sqrt{s}z)K_1(Qz) \\
  &\qquad\quad+ (e\cdot q)(q\cdot n)zK_0(Qz)
    \partial_z\left(zJ_{\Delta_V-1}(\sqrt{s}z)\right) \\
  &\qquad\quad+ (n\cdot e)\partial_z\left(QzK_1(Qz)\right)
  \partial_z\left(zJ_{\Delta_V-1}(\sqrt{s}z)\right)
    \biggr)
\end{align*}
The matrix element $M_{\mu}$ of the electromagnetic current is defined by
\begin{align}
  S_{int} \equiv ie\frac{g_{V\gamma S}}{m_V}c_0c_VR\,\delta^{(4)}(k-p_0-q)
  \,n^{\mu}M_{\mu}
  \end{align}
  with
  \begin{align*} 
  M_{\mu} &= Qe_{\mu}\int dz\,z^{\Delta_0-1}\times\left(
    (k\cdot q)z^2J_{\Delta_V-1}(\sqrt{s}z)K_1(Qz) +
    \partial_z\left(zK_1(Qz)\right)
    \partial_z\left(zJ_{\Delta_V-1}(\sqrt{s}z)\right)\right) \\
  &- Qk_{\mu}(q\cdot e)\int dz\,z^{\Delta_0-1}\,
    z^2J_{\Delta_V-1}(\sqrt{s}z)K_1(Qz) 
  + q_{\mu}(e\cdot q)\int dz\,z^{\Delta_0-1}\,zK_0(Qz)
  \partial_z\left(zJ_{\Delta_V-1}(\sqrt{s}z)\right) 
\end{align*}
and is gauge invariant by construction,
\begin{align*}
 q_{\mu}M^{\mu} = Q(e\cdot q)\int dz\,z^{\Delta_0-1}
 \partial_z\left(zJ_{\Delta_V-1}(\sqrt{s}z)\right)
 \left(\partial_z\left(zK_1(Qz)\right)+QzK_0(Qz)\right) \equiv 0\;.
\end{align*}
 Hence the matrix element $M_{\mu}$ can be written as
\begin{align}
  M_{\mu} &= Q\left(e_{\mu}(k\cdot q) - k_{\mu}(e\cdot q)\right)
  \int dz\,z^{\Delta_0+1}\, J_{\Delta_V-1}(\sqrt{s}z)K_1(Qz) \\
  &+ \left(-Q^2e_{\mu} + q_{\mu}(e\cdot q)\right)
  \int dz\,z^{\Delta_0}\,K_0(Qz)
  \partial_z\left(zJ_{\Delta_V-1}(\sqrt{s}z)\right)\;. \nonumber
\end{align}
If $\Delta_V+\Delta_0> 0$, the first  integral equals
\begin{align}
\label{I'1}
  \mathcal{I}_{1} = Q^{-\Delta_0-2}
  \left(\frac{s}{Q^2}\right)^{\frac{\Delta_V-1}{2}}
    \frac{\Gamma\left(\frac{\Delta_V+\Delta_0}{2}+1\right)
    \Gamma\left(\frac{\Delta_V+\Delta_0}{2}\right)}
  {2^{-\Delta_0}\Gamma(\Delta_V)}
  \times~ _{2}F_{1}\left(\frac{\Delta_V+\Delta_0}{2}+1,\frac{\Delta_V+\Delta_0}{2};
    \Delta_V;-\frac{s}{Q^2}\right)
\end{align}
and the second one is
\begin{align}
\label{I'0}
  \mathcal{I}_0 &= \Delta_V Q^{-\Delta_0-1}
  \left(\frac{s}{Q^2}\right)^{\frac{\Delta_V-1}{2}}
    \frac{\Gamma^2\left(\frac{\Delta_V+\Delta_0}{2}\right)}
  {2^{1-\Delta_0}\Gamma(\Delta_V)}
  \times ~ _{2}F_{1}\left(\frac{\Delta_V+\Delta_0}{2},\frac{\Delta_V+\Delta_0}{2};
    \Delta_V;-\frac{s}{Q^2}\right) \\
  &- s^{1/2}Q^{-\Delta_0-2}
  \left(\frac{s}{Q^2}\right)^{\frac{\Delta_V}{2}}
    \frac{\Gamma^2\left(\frac{\Delta_V+\Delta_0}{2}+1\right)}
  {2^{-\Delta_0}\Gamma(\Delta_V+1)}
  \times ~ _{2}F_{1}\left(\frac{\Delta_V+\Delta_0}{2}+1,\frac{\Delta_V+\Delta_0}{2}+1;
    \Delta_V+1;-\frac{s}{Q^2}\right)\; . \nonumber
\end{align}
Setting
\begin{align*}
  x = \frac{Q^2}{s+Q^2} = -\frac{1}{2}\frac{q^2}{p_0\cdot q}\,,\qquad 
  k^2 = (p_0+q)^2 = -s\,,\quad p_0^2=0\,,\quad Q^2=q\cdot q
\end{align*}
we get
\begin{align}
  \mathcal{I}_1 &= (\Delta_V+\Delta_0)
  \frac{\Gamma^2\left(\frac{\Delta_V+\Delta_0}{2}\right)}
  {2^{1-\Delta_0}\Gamma(\Delta_V)}
  \times Q^{-\Delta_0-2}x^{\frac{\Delta_0+3}{2}}
  \left(1-x\right)^{\frac{\Delta_V-1}{2}}
  ~ _{2}F_{1}\left(\frac{\Delta_V+\Delta_0}{2}+1,\frac{\Delta_V-\Delta_0}{2};
    \Delta_V;1-x\right) \\
  \mathcal{I}_0 &= \frac{\Delta_V\Gamma^2\left(\frac{\Delta_V+\Delta_0}{2}\right)}
  {2^{1-\Delta_0}\Gamma(\Delta_V)} Q^{-\Delta_0-1}
  x^{\frac{\Delta_0+1}{2}}(1-x)^{\frac{\Delta_V-1}{2}}\biggl(
 ~ _{2}F_{1}\left(\frac{\Delta_V+\Delta_0}{2},\frac{\Delta_V-\Delta_0}{2};
    \Delta_V;1-x\right)\nonumber\\
  &\qquad\quad -\frac{(\Delta_V+\Delta_0)^2}{2\Delta_V^2}\times 
  (1-x)
  ~ _{2}F_{1}\left(\frac{\Delta_V+\Delta_0}{2}+1,\frac{\Delta_V-\Delta_0}{2};
    \Delta_V+1;1-x\right)\biggr)
\end{align}
Denoting $
  \mathcal{I}_1 \equiv Q^{-\Delta_0-2}A_1(x)\,, ~~~~
  \mathcal{I}_0 \equiv Q^{-\Delta_0-1}A_0(x)\,,$
 the amplitude $M_{\mu}$ reads
\begin{align}
    M_{\mu} &= Q^{-\Delta_0-1}\biggl(
    \left(e_{\mu}(k\cdot q) - k_{\mu}(e\cdot q)\right)A_1(x)
    + \left(-Q^2e_{\mu} + q_{\mu}(e\cdot q)\right)A_0(x)\biggr)\,.
\end{align}
The vector field polarizations must be summed in the square of the
 transition amplitude. The terms linear in $e$ vanish whereas
the quadratic terms are written as
$$
  \overline{e_{\mu}e_{\nu}} = \eta_{\mu\nu} + \frac{k_{\mu}k_{\nu}}{s} ~~,~~
  \overline{e_{\mu}(e\cdot q)} = q_{\mu} + \frac{k\cdot q}{s}k_{\mu} ~~,~~
  \overline{(e\cdot q)(e\cdot q)} = q^2 + \frac{(k\cdot q)^2}{s}~~.$$
Hence, with
 $ p_0\cdot q = -yQ^2\,\equiv\frac{-Q^2}{2x}$,
 we get :
\begin{align}
  M_{\mu}M_{\nu} &= Q^{-2\Delta_0+2}
  \left(\eta_{\mu\nu} - \frac{q_{\mu}q_{\nu}}{Q^2}\right)
  \left(A_0(x)+(y-1)A_1(x)\right)^2 \nonumber\\
  &+ \frac{Q^{-2\Delta_0+2}}{s}\left(p_{0\mu}p_{0\nu} 
    + y(p_{0\mu}q_{\nu}+p_{0\nu}q_{\mu})
    + y^2q_{\mu}q_{\nu}\right)\left(A_0^2(x)+\frac{s}{Q^2}A_1^2(x)\right)\,.
\end{align}
The electromagnetic hadronic tensor thus reads
\begin{align}
  W_{\mu\nu} &= \frac{1}{2}\left(\frac{g_{V\gamma S}}{m_V}c_0c_VR\right)^2
  \sum_{n}M_{\mu}M_{\nu}\frac{1}{2E}
  \delta\left(E-\sqrt{s+\vec{\boldmath{k}}^2}\right) \\
  &= \frac{c}{2}\frac{g^2_{V\gamma S}}{M^2_VR^2}
  \Lambda^{2\Delta_0-1}\sqrt{s}
  \sum_{n}\delta(s-m_n^2)M_{\mu}M_{\nu}\,
  = \frac{c}{4\pi}\frac{g^2_{V\gamma S}}{M^2_VR^2}
  \Lambda^{2\Delta_0-2} M_{\mu}M_{\nu}\;. \nonumber
\end{align}
where $c$ is a dimensionless constant. The structure functions then read
\begin{align}
\label{VF1F2}
  F_1(x,Q^2) &= \frac{c}{4\pi}\frac{g^2_{V\gamma S}}{M^2_VR^2}
  \left(\frac{\Lambda}{Q}\right)^{2\Delta_0-2}
  \left(A_0(x)+(\frac{1}{2x}-1)A_1(x)\right)^2 \nonumber\\
  F_2(x,Q^2) &= \frac{c}{4\pi}\frac{g^2_{V\gamma S}}{M^2_VR^2}
  \left(\frac{\Lambda}{Q}\right)^{2\Delta_0-2}\frac{1}{2x}
  \left(\frac{x}{1-x}A_0^2(x)+A_1^2(x)\right) \;.
\end{align}
As may have been anticipated from dimensional arguments, the $Q^2$
dependence of the structure functions induced by vector intermediate
states is the same as for the scalar intermediate states. It is
controlled only by the conformal dimension $\Delta_0$ of the scalar
initial state and does not depend on the conformal dimension
$\Delta_V$ of the vector intermediate states. We note however that
$\Delta_V$ is not restricted by gauge invariance.

\section{\normalsize Scaling behaviour and interpretation}

We can thus repeat our argument and advocate that the case
$\Delta_0 = 1$ indeed dominates the large $Q^2$ behaviour of the DIS
cross-section. The tensorial behaviour is now sufficiently rich to
consider a meaningful correspondence between the AdS/QCD results and
real physics, namely the partonic description of the structure
functions. To do this, let us recall the reader that the cornerstone
of the interpretation of the DIS amplitude as a scattering of a hard
probe on pointlike fermions inside the hadron, is the Callan-Gross
relation \cite{CG} \beq F_{2}(x,Q^2)\;=\; 2 x F_{1}(x,Q^2) \, , \eq
which states that the longitudinal cross section is small with respect
to the transverse cross section ($\sigma_{L}/ \sigma_{T} \sim 1/Q^2$).
We note that by a superposition of scalar and vector final states the
Callan-Gross relation can now hold.

The compatibility system of equations is straightforwardly deduced
from Eqs. (\ref{PSF1F2}), (\ref{VF1F2}) and reads (after 
introducing an effective scalar coupling $C'$ and redefining in
Eq.~(\ref{VF1F2}) the constants $ K_{\Delta_V} = C\,\frac{g^2_{V\gamma
    S}}{M^2_V\,R^2} $) 
\beq
\label{CGcond}
C' \frac{A^2(x)}{2x} + \sum\limits_{\Delta_V} \frac {K_{\Delta_V}}{2x}\left[ \frac{x}{1-x}A_0^2(x) 
+ A^2_1(x)  \right] = \sum\limits_{\Delta_V} 2 x K_{\Delta_V} \left[ 
A_0(x) + \left( \frac{1}{2x}-1   \right)A_1(x)  \right]^2\;. 
\eq
A series expansion of the hypergeometric functions in
Eqs.~(\ref{I'1}), (\ref{I'0}) and some straightforward but tedious
algebra show that one can fix recursively the parameters
$K_{\Delta_V}$ to fulfill this equation.

We can now interpret our results. We are able to obtain a non empty
intersection between the QCD description of deep inelastic scattering
on a scalar hadron at medium $x$ and the simple AdS/QCD picture in the
supergravity approximation proposed by
Polchinski and Strassler, provided that
\begin{itemize} 
\item we accept some fine tuning relating the scalar $\to$ scalar and
  scalar $\to$ vector amplitudes, enabling us to mimic the scattering
  on a spinor constituent through the vanishing of the longitudinal
  structure function $F_L(x)=F_2(x)-2xF_1(x)$. 
\item when considering medium $x$ physics in the normal scaling region, we use 
the minimal value, $\Delta_0 = 1$ for the conformal  dimension of the hadronic field.
  The choice $\Delta_0 = 1$ may look - and indeed is - contradictory
  with many recent works in the framework of the AdS/CFT
  correspondence, which take for granted that the value of $\Delta $
  is to be fixed by the dimension of the interpolating current able to
  create the hadron from the vacuum. In terms of quark fields, this
  yields values such as $\Delta = 2$ for a scalar meson.
  Our point is that such an assignment is obviously completely
  incompatible with the scaling property of DIS on a meson at medium $x$, which is
  the most basic result in favor of the validity of quantum
  chromodynamics as the theory of strong interactions. We thus propose
  to take seriously the fact that the value, $\Delta_0 = 1$ describing
  a "partonic" fluctuation of the hadron is the right choice if we
  want to consider the AdS/QCD correspondence as a useful tool to
  describe strong interactions in a regime where a hard probe
  distinguishes the inner content of the hadrons. Let us note that the analysis of
   Ref. \cite{CC} shows   that in the small $x$ region, with saturation effects taken into account,
the quite different scaling   behaviour - known as geometric scaling - may result
from another choice of conformal dimension, namely $\Delta_0=2$.

\end{itemize}

\section{\normalsize  Conclusions}

By performing a sum on the final states of deep inelastic scattering
on a scalar target, we derived a sensible expression for the structure
functions in the regime of Bjorken scaling.  Our strategy has been
motivated by the well known - but quite badly understood - success of
the concept of quark-hadron duality \cite{QH}. We refer here to the "global" Bloom-Gilman
parton hadron duality which states, for instance in the analysis 
of the total cross section in electron positron annihilation, that the sum of hadronic amplitudes over the many
resonances which may be produced equals the sum of the contributions of the quark amplitudes. This does not
 preclude that "local" parton hadron duality \cite{Dok} survives at strong coupling, as has recently been shown not to be the case
 by some recent work \cite{Hatta08} in the framework of N=4 SYM. 
 
 Our calculation is of course simplistic and one should not take seriously the $x-$dependence
of the resulting structure functions.  We however think that it is
exemplary in the sense that it demonstates that the quark content of
the hadron, as it emerges from a partonic interpretation of the DIS
cross section, can be borne out of a computation where only hadrons
live in the 5-dimensional bulk.  Although the right tensorial
structure of the hadronic part is obtained at the price of some fine
tuning of the parameters controlling the relative contributions of
scalar and vector final states, this result may be viewed as a
positive sign of the validity of some AdS/QCD approach to strong
interaction physics in the domain where a partonic description has
been proven to be very effective.

\section*{Acknowledgements}

\noi We are grateful to Stanley Brodsky, Stephane Munier, Francesco
Nitti and Guy de Teramond for useful discussions and correspondence.
This work is partly supported by the French-Polish scientific
agreement Polonium, the Polish Grant N N202 249235, the ECO-NET program, contract 12584QK, and the
``consortium Physique des deux infinis''.

\section*{Appendix}

We explicitly show in this appendix that the $Q^2$-dependence of the
structure functions is the same in the soft-wall and the hard-wall
models.

Following \cite{Karch} a background dilaton field $\chi(z)$ is
introduced in the AdS metric $g_{ij}$. Then the action which describes
5d scalar fields propagating on this background reads
\begin{align}
  S = \int d^4x ~dz \sqrt{-g}e^{-\chi}\left(g^{ij}\partial_i\phi\partial_j\phi +
    m^2_S\phi^2\right)\,,
\end{align}
and the Laplacian reads
\begin{align}
  \frac{e^{\chi}}{\sqrt{-g}}\partial_i
  \left(e^{-\chi}\sqrt{-g} g^{ij}\frac{\partial \phi}{\partial x^j}\right) 
  = m_S^2\phi\,.
\end{align}
In Poincar\'e coordinates, the Laplacian becomes
\begin{align}
  z^2\square\phi + 
  z^5e^{\chi}\partial_z\left(z^{-3}e^{-\chi}\partial_z\phi\right) 
  = (m_SR)^2\phi\,.
\end{align}
Looking again for a solution that is a plane-wave in Minkowski space,
and setting
\begin{align}
  \phi(x,z) = e^{ip\cdot x}e^{\chi(z)/2}z^{3/2}\psi(z)\,,  
\end{align}
the Laplacian equation becomes a Schr\"odinger-like equation
\begin{gather}
  \frac{d^2\psi}{dz^2} - V(z)\psi = p^2\psi \\
  V(z) = \frac{(m_SR)^2+15/4}{z^2} + \frac{3}{2z}\partial_z\chi
  + \frac{1}{4}(\partial_z\chi)^2 - \frac{1}{2}\partial_z^2\chi\,,
\end{gather}
which we solve in the WKB approximation, since we are interested in the highly
excited states. The classically allowed region corresponds to
$s=-p^2>V(z)$. Let $z_0<z_{\infty}$ be the classical turning
points, $V(z_0)=V(z_{\infty})=s$, where the semi-classical
approximation breaks down. The solution is
\begin{gather}
  \psi(z) = \frac{C}{\sqrt{K(z)}}
  \sin\left(\frac{\pi}{4}+\int_z^{z_{\infty}} K(t)dt\right)\,,\qquad
  z_0<z<z_{\infty}\,, \\
  K(z) = \sqrt{s-V(z)}\,,\qquad s=-p^2\,,
\end{gather}
with the Bohr-Sommerfeld quantization rule
\begin{align}
  \int_{z_0}^{z_{\infty}} K(z)dz = \pi\left(n+\frac{1}{2}\right)\,.
\end{align}
The normalization integral and the density of states read
\begin{gather}
  C^2R^3\int_{z_0}^{z_{\infty}} dz\,\frac{e^{\chi(z)}}{K(z)}
    \sin^2\left(\frac{\pi}{4}+\int_z^{z_{\infty}} K(t)dt\right) = 1 \\
  \frac{dn}{ds} = 
  \frac{1}{2\pi}\int_{z_0}^{z_{\infty}} \frac{dz}{\sqrt{s-V(z)}}\,.
\end{gather}
In the limit $s\gg 1$, the argument of the sine is a function of $z$
which varies much faster than the other functions in the
integrand. Hence one can replace in this
limit the sine square by its average value $1/2$. Then we get an
approximate expression for the normalization constant,
\begin{align}
  C^{-2} \approx \frac{R^3}{2}
  \int_{z_0}^{z_{\infty}} dz\,\frac{e^{\chi(z)}}{K(z)}\,.
\end{align}
We can calculate the integrals approximately in the limit $s\gg 1$.
Indeed inside the classically allowed regions, far from the turning
points, we have then $s\gg V(z)$ and we can expand the integrand with
respect to $s$. Hence
\begin{align}
  \label{Dsoft}
  \frac{dn}{ds} &\approx \frac{1}{2\pi\sqrt{s}}
  \int_{z_0}^{z_{\infty}}dz\left(1 + \mathcal{O}\left(\frac{1}{s}\right)\right)
  \approx \frac{1}{2\pi\Lambda\sqrt{s}}\quad,\qquad
  z_{\infty}=\frac{1}{\Lambda} \\
  \label{Csoft}
  C^{-2} &\approx \frac{R^3}{2\sqrt{s}}\int_{z_0}^{z_{\infty}} dz\,e^{\chi(z)}
  \left(1 + \mathcal{O}\left(\frac{1}{s}\right)\right)
  \approx \frac{R^3}{2\sqrt{s}}\int_{z_0}^{z_{\infty}} dz\,e^{\chi(z)}
  = c_{\chi}\frac{R^3}{2\Lambda\sqrt{s}}
\end{align}
where $c_{\chi}$ is a dimensionless constant. We find that the
densities of states \eqref{Dsoft} and \eqref{Dhard}, and the
normalizations \eqref{Csoft} and \eqref{Chard}, have indeed the same
dimensional structure in the soft-wall and hard-wall models. This
generalizes the result of \cite{BBB} in the specific soft-wall model
with linear Regge trajectories.


\begin{thebibliography}{99}

\bibitem{PS1} 
J.~Polchinski and M.~J.~Strassler,
  Phys.\ Rev.\ Lett.\  {\bf 88} (2002) 031601;
O.~Andreev,
   Phys.\ Rev.\  D {\bf 67}, 046001 (2003).
  
\bibitem{BdT}
  S.~J.~Brodsky and G.~F.~de Teramond,
  Phys.\ Rev.\ D {\bf 77}, 056007 (2008);
  Phys.\ Rev.\ Lett.\  {\bf 96} (2006) 201601;
  Phys.\ Rev.\ Lett.\  {\bf 94} (2005) 201601;
  Phys.\ Lett.\  B {\bf 582} (2004) 211.
 
 \bibitem{Rad}
  H.~R.~Grigoryan and A.~V.~Radyushkin,
  arXiv:0803.1143 [hep-ph],
  Phys.\ Rev.\  D {\bf 76} (2007) 115007,
  Phys.\ Rev.\  D {\bf 76} (2007) 095007;
  Phys.\ Lett.\  B {\bf 650} (2007) 421;
  Z.~Abidin and C.~E.~Carlson,
  arXiv:0801.3839 [hep-ph].

\bibitem{Karch}
  A.~Karch {\em et.al.},
  Phys.\ Rev.\  D {\bf 74}, 015005 (2006).

\bibitem{BBB}
  C.~A.~Ballon Bayona, H.~Boschi-Filho and N.~R.~F.~Braga,
  JHEP {\bf 0803}, 064 (2008).
   
\bibitem{others}
  A.~Karch, E.~Katz,
  JHEP {\bf 0206}, 043 (2002); 
  C.~A.~Ballon Bayona, H.~Boschi-Filho and N.~R.~F.~Braga,
  arXiv:0712.3530 [hep-th];
   Y.~Hatta, E.~Iancu and A.~H.~Mueller,
  JHEP {\bf 0801}, 026 (2008).
 
 
 \bibitem{CC}
  L.~Cornalba and M.~S.~Costa,
  arXiv:0804.1562 [hep-ph].

\bibitem{PS2} 
J.~Polchinski and M.~J.~Strassler,
  JHEP {\bf 0305} (2003) 012. 
   
  \bibitem{CG}
  C.G. Callan and D.J. Gross, Phys.Rev.Lett. {\bf 22}, 151 (1984).

\bibitem{QH}
E.~D.~Bloom and F.~J.~Gilman,
  Phys.\ Rev.\ Lett.\  {\bf 25}, 1140 (1970);
   W.~Melnitchouk, R.~Ent and C.~Keppel,
  Phys.\ Rept.\  {\bf 406}, 127 (2005);
N.~Isgur {\it et al.},
  Phys.\ Rev.\  D {\bf 64}, 054005 (2001).
  
  \bibitem{Feynman}
  R.P. Feynman, Photon-Hadron Interactions, W.A. Benjamin, 1972.
  
  \bibitem{KW}
  I.~R.~Klebanov and E.~Witten,
  Nucl.\ Phys.\  B {\bf 556}, 89 (1999).
   
   \bibitem{Dok} 
 Y.~L.~Dokshitzer, V.~A.~Khoze, S.~I.~Troian and A.~H.~Mueller,
  Rev.\ Mod.\ Phys.\  {\bf 60}, 373 (1988).
  
   \bibitem{Hatta08}
   Y.~Hatta and T.~Matsuo,
  arXiv:0804.4733 [hep-th] and  arXiv:0807.0098 [hep-ph].
\end{thebibliography}
\end{document}